\pdfoutput=1
\documentclass[useAMS,usenatbib]{mnras}
\usepackage[T1]{fontenc}
\usepackage{ae,aecompl}
\usepackage{graphicx} 
\usepackage{dcolumn}  
\usepackage{bm}       
\usepackage{amssymb,amsmath}
\usepackage{color,colortbl}
\usepackage{float}
\usepackage{nicefrac}
\usepackage{xspace}
\usepackage[inline]{enumitem}

\newcommand{\etal}{{\it et al.\,}}
\newcommand{\cola}{{\tt{COLA}}\xspace}
\newcommand{\icecola}{{\tt{ICE-COLA}}\xspace}

\def\Msun{\, h^{-1} \, {\rm M_{\odot}}}
\newcommand{\hompc}{\,h\,{\rm Mpc}^{-1}}
\newcommand{\mpcoh}{\,h^{-1}\,{\rm Mpc}}

\begin{document}

\title[ICE-COLA: fast WL simulations]
{ICE-COLA: fast simulations for weak lensing observables}
\author[Izard \etal]{
Albert Izard$^{1,2,3}$, Pablo Fosalba$^1$, Martin Crocce$^1$ \\
$^1$Institut de Ci\`encies de l'Espai, IEEC-CSIC, Campus UAB, Carrer de Can Magrans, s/n,  08193 Bellaterra, Barcelona, Spain \\
$^2$Jet Propulsion Laboratory, California Institute of Technology, 4800 Oak Grove Drive, Pasadena, CA 91109, USA \\
$^3$Department of Physics and Astronomy, University of California, Riverside, CA 92521, USA
}

\twocolumn
\maketitle

\begin{abstract}
{Approximate methods to full N-body simulations
provide a fast and accurate solution to the development of mock catalogues for the modeling of galaxy clustering observables.
In this paper we extend \icecola, based on an optimized implementation of the approximate \cola method,
to produce weak lensing maps and halo catalogues in the light cone using an integrated and self consistent approach.
We show that despite the approximate dynamics, the catalogues thus produced enable an accurate modeling of weak lensing observables one decade beyond the characteristic scale where the growth becomes non-linear. In particular, we compare \icecola to the MICE-GC N-body simulation for some fiducial cases representative of upcoming surveys and find that, for sources at redshift $z=1$, their convergence power spectra agree to within one percent up to high multipoles (i.e.,  of order $1000$). The corresponding shear two point functions, $\xi_{+}$ and $\xi_{-}$, yield similar accuracy down to $2$ and $20$ arcmin respectively, while tangential shear around a $z=0.5$ lens sample is accurate down to $4$ arcmin. We show that such accuracy is stable against an increased angular resolution of the weak lensing maps. Hence, this opens the possibility of using approximate methods for the joint modeling of galaxy clustering and weak lensing observables and their covariance in ongoing and future galaxy surveys.}
\end{abstract}


\section{Introduction}

The images of distant sources are distorted as their photons travel across the gravitational field of dark-matter structures along the line of sight of observers today. The strength of the phenomenon depends on the expansion history and the growth of structure (for reviews see \citealt{Bartelmann01,Kilbinger15}). Precise measurements of this effect can tightly constrain cosmological models in a complementary way to other traditional techniques (see e.g. \citealt{Weinberg13}), and elucidate the properties of the accelerated expansion and the large-scale structure formation process, which are directly related to the properties of dark energy and dark matter, respectively. Present and planned galaxy surveys such as
the Dark Energy Survey (DES)\footnote{\url{http://www.darkenergysurvey.org/}},
the Kilo-Degree Survey (KiDS)\footnote{\url{http://kids.strw.leidenuniv.nl}},
the Hyper Suprime-Cam Subaru Strategic Program (HSC-SSP)\footnote{\url{http://hsc.mtk.nao.ac.jp}},
Euclid\footnote{\url{http://www.euclid-ec.org/}}, and
the Wide Field InfraRed Survey Telescope (WFIRST)\footnote{\url{https://wfirst.gsfc.nasa.gov}},
the Large Synoptic Survey Telescope (LSST)\footnote{\url{http://www.lsst.org}},
represent a large increase in the volume sampled by weak gravitational lensing experiments, which results in a tremendous narrowing down of the statistical uncertainties associated with the measurements. This has to go along with an equivalent reduction of systematic errors, otherwise the constraining power on cosmological models will not scale accordingly.

The process of converting the data acquired by experiments into catalogues ready for scientific usage involves many calibration and analysis steps. It is a non-linear process in which many systematic effects may be introduced, from instrumental noise and limitations to atmospheric variations \citep{Leistedt15} or astrophysical sources of errors (e.g., the confusion of objects in the same region of the sky, see \citealt{Hartlap11}). The science analysis is also prone to introduce biases if the physical properties of the sample observed (such as the signal measured and its covariance matrix) are not well understood and modeled accurately \citep{Dodelson13,Taylor13,Blot15,Sellentin16,Sellentin17}. In recent years, mock galaxy surveys have become an essential tool to model and mitigate
systematics, as well as optimizing analysis pipelines (e.g., in the Baryon Acoustic Oscillations reconstruction, see \citealt{Takahashi09,Manera13,Kazin14,Ross15}).

Synthetic data has been traditionally produced by means of $N$-body simulations, which require huge computational resources in order to meet the resolution and volumes sampled by galaxy surveys (\citealt{Kim11,Angulo12,Alimi12,Skillman14,Heitmann14,MICEI}; for a review see \citealt{Kuhlen12}). In this context,  weak lensing can be accurately modeled in simulations via ray-tracing techniques, which follow the ray propagation from the source to the observer along the perturbed path (see e.g.
\citealt{Blandford86,Jain00,Das08,Teyssier09,Li11}).
This involves intensive computations because the deflection angle needs to be constantly updated as the ray travels in order to determine the geometry at each encounter of the multiple-lens system. However, under the Born approximation, which we use in this work, integrations are much faster since photons propagate along the unperturbed path across each of the (uncorrelated) lens planes (see \citealt{White00,Fosalba08,Kiessling11,Borzyszkowski17}). This removes higher-order contributions and the coupling of lenses at different distances \citep{Krause10}, and amount to sub-percent corrections for the relevant scales and redshifts sampled by planned surveys and thus can be safely neglected in their modeling (see e.g, \citealt{Petri16}).

Large ensembles of realizations are necessary for a precise determination of covariance matrices and the optimization of some pipelines and this becomes computationally prohibitive using conventional numerical simulations. This problem has been overcome thanks to the emergence of techniques known as approximate methods, that provide an optimal solution in the compromise between accuracy and speed. The latter comes from avoiding solving the highly non-linear dynamics in virialized regions, that are expensive to resolve. Using an approximate gravity solver to evolve the dark matter density field, galaxies or haloes are modeled either using a biasing prescription (\citealt{Coles91,White13,Chuang15b,Kitaura15}) or by identifying collapsed regions in the density field \citep{Monaco02,Scoccimarro02,Tassev13,Manera15,Munari16}. For a comparison of the performance of some of these methodologies see
\cite{Chuang15a} and for a recent review see \cite{Monaco16}.

Among the suite of approximate methods recently developed, \cola \citep{Tassev13,Tassev15} stands out as a simple yet powerful alternative to exact N-body simulations. In particular, it accurately solves mildly non-linear scales by integrating numerically the equations of motion of particles in a frame comoving with respect to Lagrangian Perturbation Theory (LPT) trajectories. By using broad time steps and solving the gravitational forces using a fine Particle-Mesh (PM) algorithm, the spatial distribution of dark-matter halos can be faithfully captured whereas the computational cost is reduced by 2-3 orders of magnitude with respect to standard $N$-body methods. In \cite{Izard16} an optimal configuration of the method, named \icecola, was presented for the production of accurate mock halo catalogues adapting the parallel version of \cola developed in \cite{Koda15}. In this paper we extend the \icecola code to produce all-sky light cone catalogues on-the-fly\footnote{At run time of the simulation.}, what allows for the fast and accurate modeling of weak lensing observables.
Other recent efforts involving \cola have been developed to extend the capabilities of the original code, including Modified Gravity solvers and massive neutrinos \citep{Howlett15b,Valogiannis16,Winther17,Wright17} or relativistic corrections \citep{Borzyszkowski17}.
Plain PM simulations have also been useful to produce mock catalogs \citep{Merz05,White10}, as well as variations of the algorithm using few time steps \citep{White13,Feng16}.

Fast modeling of weak lensing is a specially challenging task, since the gravitational evolution of the relevant scales is well within the non-linear regime of structure formation. At the same time, it entails probing large volumes extending far in the observer past light cone, up to the most distant sources (e.g, the last scattering surface for CMB lensing). In this paper, we discuss the capabilities and limitations of the \icecola method for light cone observables, and show that it can meet these two basic requirements. To do so, we implement the production of matter and halo light cone catalogues on-the-fly.
The dark matter density field is then projected into two-dimensional sky maps, from which we can construct weak lensing maps in the Born approximation, following the procedure presented in \cite{Fosalba08} and \cite{MICEIII}.
Halo catalogues in the light cone are stored and can then be used to produce galaxy mocks based on the Halo Occupation Distribution approach \citep{Carretero15,MICEII, MICEIII}. Weak lensing properties are then imprinted into the halos (or galaxies) interpolating them from the weak lensing maps.

Other fast methods to produce weak lensing maps have been recently proposed in the literature. \cite{Giocoli17} is based on a halo model formalism (see references therein too) and \cite{Yu16} use a gaussianization technique. These works differ to what is presented here in that they are not self consistently built upon an N-body solver of the gravitational dynamics.

This paper is organized as follows: Section \ref{sec:lc_construction} presents the light cone construction, its outputs and post-processing pipeline. Section \ref{sec:simulations} describes the \icecola and MICE N-body simulations used for the analysis. The methodology for computing all-sky weak lensing observables and their analysis is discussed in section \ref{sec:wl_simulations},  whereas the corresponding results for the light cone halo catalogues is the subject of section \ref{sec:halo_weak_lensing_mocks}. We conclude summarizing our main results in section \ref{sec:conclusions}.

\section{All-sky light cone construction}
\label{sec:lc_construction}

Sources in the Universe are seen by observers today as they were at the time photons from the source were emitted. The time elapsed corresponds to what it takes photons to travel from the source to the observer. This ``light cone'' geometry can be implemented in a simulation by computing the light cone crossing time of each particle, determined by the intersection of its trajectory with the surface of a sphere that is centered at the observer and its physical radius shrinks at the speed of light (reaching zero at the present time).

In an $N$-body simulation, particle positions are sampled in a collection of discrete time steps. In the so-called leapfrog integration, each time step contains a pair of drift and kick operators that integrate the equations of motions and update the positions and velocities of particles respectively. The two operators are interleaved, that is, they are separated in time by half the size of the time step (see \citealt{Quinn97} and \citealt{Dehnen11} for a more detailed description of the leapfrog integration). We build the particle light cone output as the set of interpolated particle coordinates determined at their light cone crossing time. In such way, the time evolves smoothly in the radial coordinate and there are no jumps in the catalogues, as may occur in light cones built from stitching snapshots. In \icecola, the light cone routine is called after each drift and kick operators, producing a shell with the geometry sketched in Fig. \ref{fig:lc_diagram_crossing_time}. The algorithm takes the following steps:

\begin{figure}
\includegraphics[trim=0 0 3.7cm 0, width=0.99\columnwidth]{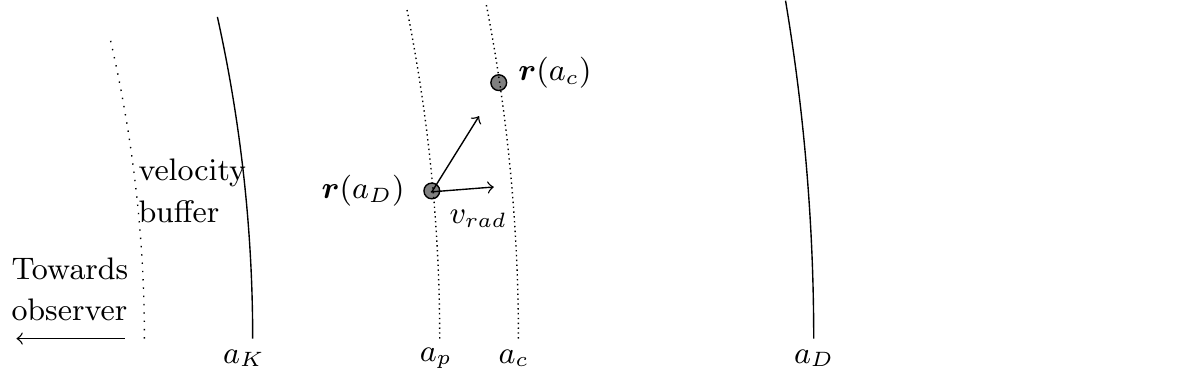}
\caption{Diagram of a particle at its crossing time. Initially, at drift time, the light cone is at position $t_D$ and the particle at $\bm{r}(t_D)$. Knowing these two quantities, together with the radial velocity, is enough to evaluate equation \ref{eq:crossing_time} that gives the crossing time $t_c$. Finally, coordinates for the particles can be interpolated at that time. Note that the velocity buffer is only necessary around $t_K$.}
\label{fig:lc_diagram_crossing_time}
\end{figure}

\begin{itemize}
 \item The time of the last drift and kick operators to evolve particle positions and velocities ($t_D$ and $t_K$ respectively) determine the radius of the shell being constructed at a given time of the simulation\footnote{Particle coordinates at intermediate time values between the last drift and kick operator can be interpolated using an extra pair of drift and kick operators.}.
 \item The shell is enlarged with a velocity buffer zone around $t_K$ to consider moving particles that may enter into the shell during the interpolation process\footnote{There is no buffer necessary around $t_D$ since at that region the crossing time is close to the time of the drift operator and therefore the displacements during the interpolation tend to zero.}.
 \item Particles that do not belong to the volume of the shell, including the buffer zone, are discarded.
 \item Given the position and radial velocity of each particle, its crossing time is computed as:

 \begin{equation}
 	t_c = \frac{c}{c+v_{rad}}(t_p-t_D)+t_D,
 \label{eq:crossing_time}
 \end{equation}

where $t_p$ is the time when the light cone boundary is at the distance of the particle position at the last drift operator.
 \item Particle coordinates are interpolated at crossing time and checked whether they are within the volume of the shell (without the buffer zones now).
 \item The particles satisfying the previous conditions are copied to the light cone catalogue. The process may be repeated for different box replicas, in which the coordinates are shifted by the box side length around the observer, as we discuss later in this section.
\end{itemize}

The light cone routine is called after each pair of kick and drift operators. To make it more efficient, we pre-compute and tabulate all the time dependent functions and do an interpolation when these have to be evaluated. This is the case for conversions between cosmological distances and times and the time-varying quantities in the temporal operators that interpolate particle coordinates to their crossing times. We have estimated that with as few as 10 time samples within each time-step the accuracy achieved is sufficient, since they are smoothly-varying quantities. In particular, we have checked that the errors in particle coordinates arising from these approximations are much lower than the spatial resolution of the simulation.

Given that the simulated comoving volume has periodic boundary conditions, it makes adjacent box replicas used to construct the full light cone around the observer have a continuous density field distribution across the entire light cone volume. We take advantage of this to produce light cones sampling larger volumes (allowing to produce catalogues that extend to higher redshifts and cover more sky area). The disadvantage of this approach is that repeated structures along the line of sight could potentially appear in the catalogue. However, they will be separated a distance given by the box size and they have typically crossed the light cone boundary at a different cosmological time, since their distance to the observer will likely be different. Therefore, no obvious repetitions of structures will be observed in the resulting light cone volume (see also \citealt{Fosalba08, MICEIII}).

Writing light cone catalogs containing the particle information results in huge data volumes, that demand not only large storage quotas but also require large computing time to be processed because of the limited input/output bandwidth. In the scope of a fast method, it is desirable to generate high-level processed catalogs that can be post-processed more easily. We describe next the two types of catalogs we generated on-the-fly.

\subsection{On-the-fly projected matter density fields}
\label{sec:hp_maps}

We implement and adapt to the particularities of \cola the ``onion universe'' method originally developed by \cite{Fosalba08} to generate compressed outputs of the matter density field in the light cone that are used to derive weak lensing maps in post-processing. The particle distribution in the light cone is interpolated into a three-dimensional grid using spherical coordinates around the observer. The radial coordinate is binned into several thin shells and the sky area is discretized into two-dimensional pixels using the Healpix software \citep{Gorski05}. In that way, the volume is subdivided into several slices of Healpix maps, whose values give the dark-matter particle count
in a given pixel.

For the ease of comparison with MICE-GC reference N-body simulation, we follow \cite{MICEI} and use 265 shells in the redshift range $0<z<1.4$ having  a width of $\sim7 (18) \mpcoh$ at low (high) redshift, corresponding to steps of $\sim35\, {\rm Myears}$ in look-back time. It is left for future work the study of how many radial bins are enough to recover accurately the matter field within the accuracy of \cola. The Healpix maps have an angular resolution given by the parameter  $N_{side}=2048$, which produces $12N_{side}^2 \approx 50$ million pixels of $\theta \simeq 1.7\, {\rm arcmin}$ length. The data volume for storing all the maps is 50 GB, which represents a compression factor of two orders of magnitude with respect to the whole particle information generated by the light cone output.

\subsection{On-the-fly halo catalogues}
\label{sec:fof_lc}

The \cola version of \cite{Koda15}
includes a parallel Friends-of-Friends (FoF) algorithm \citep{Davis85} that runs on-the-fly on particle snapshots, and it is based on the publicly available serial code from the $N$-body shop of the University of Washington\footnote{\url{http://www-hpcc.astro.washington.edu/tools/fof.html}}. In \icecola we adapted it to produce FoF catalogues in the light cone.

A halo finder explores the environment around particles and links those that belong to the same halo. In a comoving snaphsot, the environment of regions near the edges of the simulated box can be completed with the particles close to the opposite side of the box, thanks to the periodic boundary conditions. However, this is no longer possible when the algorithm is run in the light cone particle distribution, since the crossing time varies across the volume and thus differs for each box replica. Therefore, at the edges of the box it is needed the information of the neighboring replica, which has to be computed using the adequate crossing times. That is, the environment of a particle close to the box edge has to be completed with the volume corresponding to the neighboring replica. But since only one replica is processed at a time, it is necessary to add buffer zones around box edges and provide in that way the required spatial environment to the algorithm.

Besides, each call to the light cone routine builds a spherical shell of the light cone volume. Therefore, buffer zones are also needed around the spherical shell boundaries. In summary, when FoF catalogues are requested, the light cone algorithm builds a subvolume containing the shell and the buffer zones\footnote{Note that it shall add an additional buffer for the velocities, as described before.}. Then the FoF code is run in the corresponding subsample of particles and finally only the halos whose center of mass lay within a given spherical shell and the local box replica are added to the catalogue.

\begin{figure} 
\includegraphics[width=0.99\columnwidth]{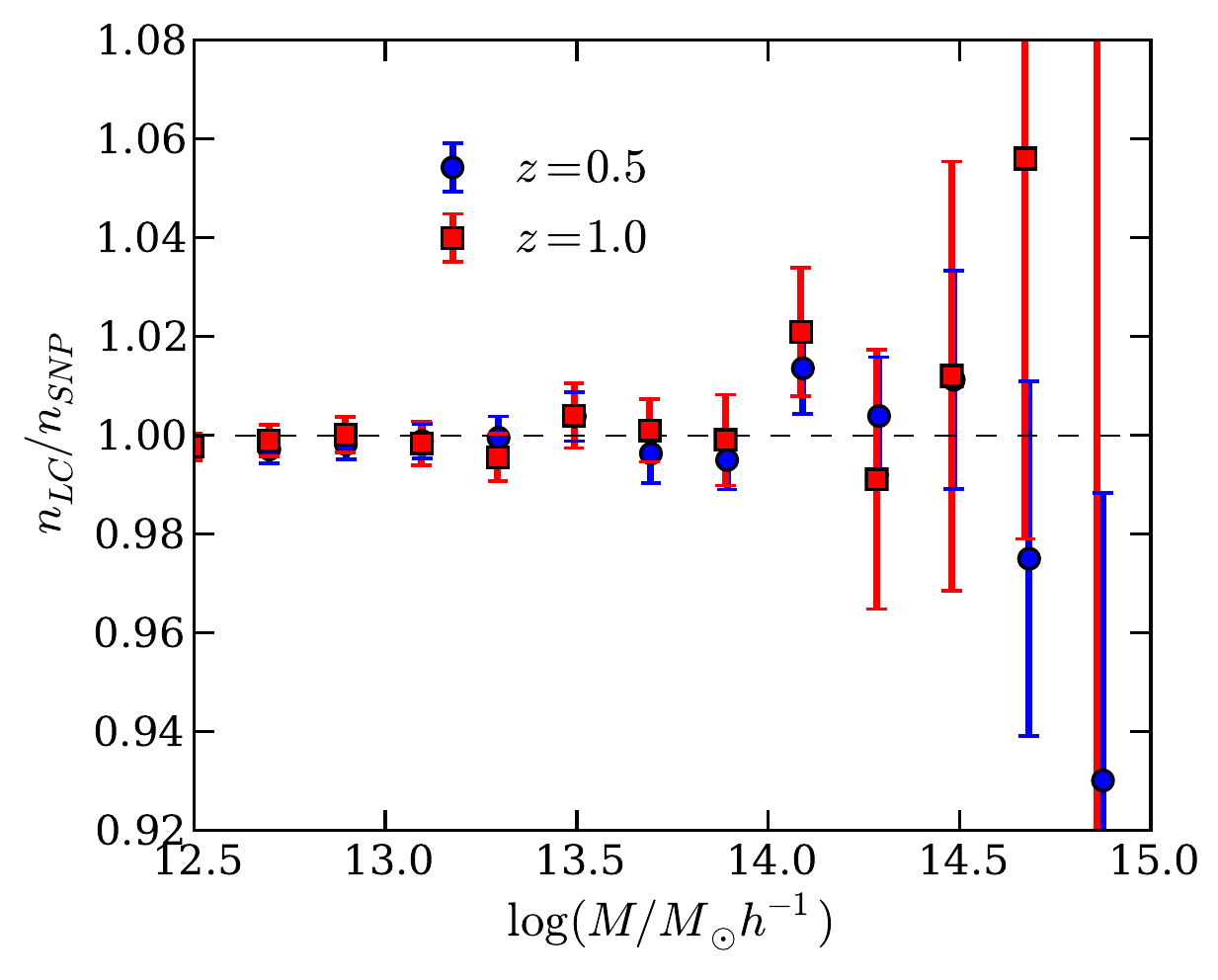}
\caption{Comparison of the mass function measured in the light cone and in the snapshots. Error bars display Jack-Knife estimates from the measurements of the comoving catalogue and we assume that the measurements in the light cone have the same error. Differences are very small or within the error bars and thus there are no signs of errors in the algorithm for generating FoF in the light cone.}
\label{fig:mf_lc_vs_snp}
\end{figure} 

An incorrect implementation of buffer zones could lead to an incorrect detection of haloes, especially with the largest ones, that would be fragmented into small pieces. We set the width of the buffer zones to $7\mpcoh$. We try to check for the fragmentation effect by comparing the mass function in the light cone and in snapshots at redshift 0.5 and 1. We make a subsample of halos in a spherical shell centered at these redshifts and having a volume equal to the volume of the box (this can be achieved by choosing the appropriate width of the shell). In this way we expect that the measurements have similar error bars. In Fig. \ref{fig:mf_lc_vs_snp} we show the ratio of the mass functions measured in the light cone and in the snapshot. Differences are well within the percent level at low masses and compatible with error bars at high masses. We also check that varying the buffer size does not modify our results.
 We thus conclude from this analysis that there are no signs of systematic biases in the algorithm for generating FoF in the light cone.

\subsection{Post-processing pipeline}
\label{sec:pipeline}

The dark-matter light cone catalogues presented in this section are generated on-the-fly and ease considerably the post-processing steps needed to produce science ready catalogues. However, an additional processing step is required to produce weak lensing observables. One has to transform dark-matter counts maps into weak lensing ones and combine them with the light cone halo catalogues. We do this in a post-processing pipeline that we briefly describe below\footnote{For these computations we employ the Healpy software tools, \url{https://healpy.readthedocs.io/en/latest/}} and that follows the approach presented in \cite{Fosalba08} (see also \citealt{MICEIII}). For further details and the assumptions involved we refer the reader to  Sec. \ref{sec:wl_simulations} and \ref{sec:halo_weak_lensing_mocks}, where results are also discussed. The basic steps of the weak-lensing post-processing pipeline are:

\begin{itemize}
\item The convergence field is determined in the Born approximation by integrating along the line-of-sight the 2D dark matter counts maps weighted by the lensing kernel (eq. \ref{eq:convergence_discrete}).
\item Next, we follow \cite{Hu00} and use a simple all-sky relation to transform the convergence into the shear field in harmonic space. The E-mode shear is then given by eq. \ref{eq:shear_e_mode}.
\item We transform back to angular space to obtain the shear map. Note that these transformations require the weak lensing maps to be all-sky to avoid boundary effects. All these steps are repeated for each of the thin shells that sample the whole redshift range $0 < z < 1.4$, having at the end a suite of 2D maps (the ``onion universe'' as introduced in \citealt{Fosalba08}) of the weak lensing effect based on the Healpix discretization.
\item Finally, halos are assigned convergence and shear values. To do so, we determine the pixel in the 2D map, for a given redshift shell, that corresponds to the halo position and assign its weak lensing quantities to the halo.

\end{itemize}

\section{Simulations}
\label{sec:simulations}

In this section we describe the numerical simulations used in this paper to validate the performance of \icecola for the modeling of weak-lensing observables.

\subsection{\icecola simulations}
\label{sec:icecola_simulations}

In our previous work \cite{Izard16} we investigated the impact of various parameters of the \cola method on the matter power spectrum, the halo abundance and clustering. We showed that, in order to capture halo formation accurately at a given output time, more than 10 simulation time-steps need to be computed from the initial conditions to that given output time. We suggested a configuration using 40 time steps linearly distributed with the scale factor between an initial redshift of $z_i=19$ and $z=0$ and using a cell size for the computation of forces that is 3 times smaller than the mean inter-particle distance separation. Simulations using this configuration predict the mass function with a $5\%$ accuracy and the matter power spectrum within the percent level at scales $k\lesssim1\hompc$. In this paper we use this optimal configuration for runs with $2048^3$ particles and a box size of $1536\mpcoh$, which results in a particle mass of $2.93\times10^{10}\Msun$. Both the particle mass, the cosmological model and the input linear power spectrum are identical to the reference MICE-GC N-body simulation (see Sec. \ref{sec:micegc}).

For this paper we make use of the capabilities of \icecola to generate an all-sky light cone on-the-fly. The simulated box is replicated twice in each cartesian direction\footnote{This generates 64 box replicas in total.}, allowing to build a light- cone that samples the redshift interval $0<z<1.4$. From the full particle information, two derived catalogues are produced and stored: two-dimensional matter density maps projected on the sky (see Sec. \ref{sec:hp_maps}) and Friends-of-Friends (FoF) halo catalogue \citep{Davis85} with a linking length of 0.2 (see Sec. \ref{sec:fof_lc}). The all-sky FoF catalogue include objects with 50 or more particles, generating 15 Gb of data. This data volume is to be added to the 50 Gb of the particle dark-matter light cone outputs in Healpix maps. The simulation was run on 1024 cores in 100 minutes, where roughly 50 $\%$ of the time was spent running the FoF algorithm in the light cone volume covering the 64 box replicas.

Our default Healpix resolution for the weak lensing maps is $N_{side}=2048$. However we produced a higher resolution version of them to test in Sec. \ref{sec:halo_weak_lensing_mocks} whether the accuracy of the method is limited by the approximated dynamics of \cola or by the pixel scale employed. These maps have $N_{side}=4096$, which correspond to $\theta \simeq 0.85\, {\rm arcmin}$ pixels, and cover only one octant of the sky to reduce the total data volume. We generated them at post-processing from a light cone catalogue containing three-dimensional particle information, which we projected at post-processing into two-dimensional maps as explained in Sec. \ref{sec:hp_maps}.

\subsection{MICE-GC: the benchmark $N$-body run}
\label{sec:micegc}

Our benchmark $N$-body simulation is the MICE Grand Challenge (MICE-GC)\footnote{More information is available at \texttt{http://www.ice.cat/mice}.}. It evolved $4096^3$ particles in a volume of $(3072\mpcoh)^3$ using the \textsc{gadget-2} code \citep{Springel05} assuming a flat $\Lambda$CDM cosmology with $\Omega_m=0.25$, $\Omega_{\Lambda}=0.75$, $\Omega_b=0.044$, $n_s=0.95$, $\sigma_8=0.8$ and $h=0.7$. This results in a particle mass of $2.93\times10^{10}\Msun$. The initial conditions were generated at $z_i=100$ using the Zel'dovich approximation and a linear power spectrum generated with \textsc{camb}\footnote{\texttt{http://camb.info}}. Haloes were identified using a the FoF algorithm \citep{Davis85} with a linking length of 0.2.
Note that MICE-GC and the \icecola runs have different box sizes and therefore the comparison between them will be affected by sample variance.

This simulation and its products have been extensively validated. Of special interest in this work is the modeling of weak gravitational lensing, presented in \citet{MICEIII}. The dark-matter and halo outputs are described in \citet{MICEI} and \citet{MICEII}. The Halo Occupation Distribution implementation used to produce galaxy mocks is detailed in \citet{Carretero15} while \citet{Hoffmann15b} focuses on the higher-order clustering. In this work we make use of the weak lensing maps and halo catalogues in its past light cone.

The weak lensing maps in MICE-GC employ a finer pixel size of $\theta \simeq 0.43\, {\rm arcmin}$ (i.e, Healpix $N_{side}=8192$), meaning that they have a better angular resolution than the catalogues of \icecola, but we have checked that degrading to the same resolution than \icecola does not change results down to $\simeq 0.5\, {\rm arcmin}$. As for the light cone outputs, the convergence and matter density maps are all-sky, whereas the corresponding MICE-GC halo catalogues are produced only for one octant of the sky.

\section{Modeling weak lensing}
\label{sec:wl_simulations}

\begin{figure}
\includegraphics[width=0.99\columnwidth]{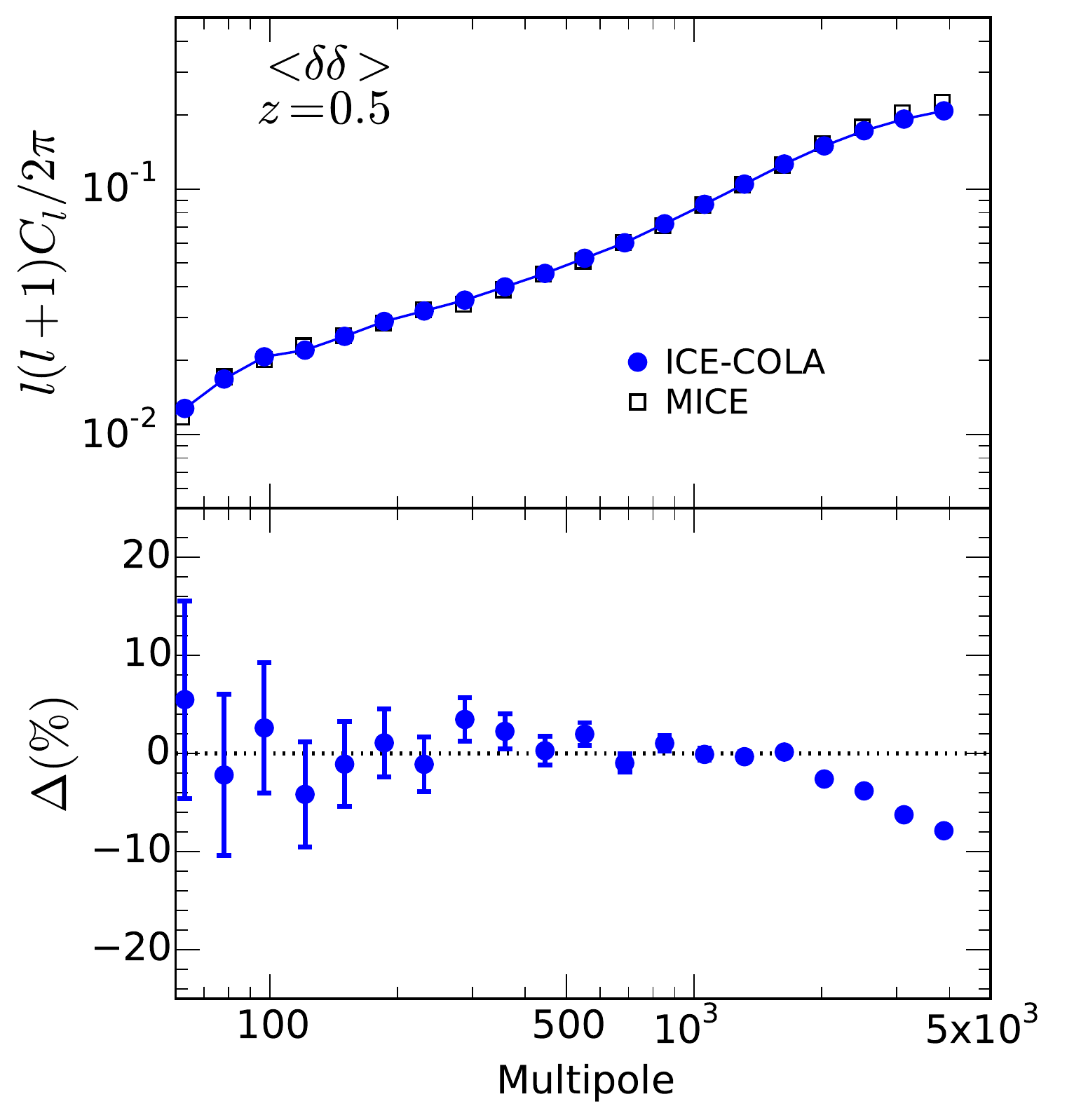}
\caption{
Matter angular power spectrum for a slice at redshift $z=0.5$ and a width of $\Delta z=0.1$\protect\footnotemark. The lower panel shows the relative differences between \cola and MICE-GC, that are within the error bars up to $l\sim2000$ and within $10\%$ up to $l=4000$.}
\label{fig:cl_dd_m}
\end{figure}

As mentioned in Sec. \ref{sec:hp_maps}, we make use of the two-dimensional projected matter density field generated on-the-fly to compute weak lensing maps in the Born approximation. The resulting convergence maps receive contributions from the intervening matter fluctuations  weighted by the so-called lensing efficiency, which peaks around half the distance to the source. Throughout this paper we choose a particular configuration to study weak lensing, in which the sources are centered at redshift 1 and the lenses at redshift 0.5.. It is thus informative to investigate the accuracy in modeling the dark-matter clustering at the latter distance to assess the accuracy of \icecola to model weak lensing observables. To illustrate this, in Fig. \ref{fig:cl_dd_m} we compare the angular matter power spectrum at $z =0.5$ for \icecola (circles) against MICE-GC (squares), where we have subtracted the shot-noise contribution to the power spectrum according to

\footnotetext{ Note that this width is unrelated to the width used to construct the 2D maps. }

\begin{equation}
	C_l^{noise} = f_{sky}\frac{4\pi}{N},
	\label{eq:shot_noise_cl}
\end{equation}

\noindent
where $N$ is the number of objects in the sample and $f_{sky}$ is the fractional area of the sky of the catalogue (1 if it is all-sky). At small scales it represents a 1\% correction for \cola .
We display Gaussian error bars, calculated by considering the number of modes sampled at each multipole $l$-bin

\begin{equation}
	\sigma(C_l) = \sqrt{ \frac{2}{f_{sky}\Delta l (2l+1)} } (C_l + C_l^{noise}),
\end{equation}

\noindent
where $\Delta l$ is the multipole bin width. In the lower panel displaying the ratio, we add the respective error-bars in quadrature.

We find that the agreement between \icecola and MICE-GC is very good up to $l=2000$ and within $10\%$ at $l=4000$. At $z=0.5$ and in the flat-sky limit (which should be very accurate for the relevant scales) these multipoles project onto  $k\sim l/r\sim1.5\hompc$ and $3\hompc$ respectively, where we have used that $r(z=0.5)=1346\mpcoh$ for the cosmology assumed. We note that, at these scales, most of the contribution to the power spectrum comes from the internal halo structure \citep{vanDaalen15}, and therefore we expect a decline in the power for \cola since dynamics within haloes are not fully resolved (see also the matter power spectrum of the same simulation in Fig. 8 of \citealt{Izard16}).

\subsection{Convergence maps}
\label{sec:kappa_maps}

The convergence field characterizes the fractional change in the image size. In the Born approximation, the contribution from all the lens planes between the observer and the sources at a distance $\chi$ can be calculated by an integral:

\begin{equation}
	\kappa(\bm\theta,\chi) = \frac{3H_0^2\Omega_m}{2c^2} \int_0^\chi \delta(\bm\theta,\chi')\frac{(\chi-\chi')\chi'}{a\chi}\mathop{d\chi'}
	\label{eq:convergence_born}
\end{equation}

\noindent
where $H_0$ is the Hubble constant, $\Omega_m$ is the matter density parameter, $c$ is the speed of light, $a$ is the scale factor, and $\delta(\bm\theta,\chi)$ is the 3D matter density contrast at the angular position $\bm\theta$ and radial distance $\chi$. Following \cite{Fosalba08}, we compute the convergence maps by discretizing the equation in several radial bins and angular pixels as

\begin{equation}
	\kappa(\bm\theta_i,\chi) = \frac{3H_0^2\Omega_m}{2c^2} \sum_j^{\chi_j<\chi} \delta(\bm\theta_i,\chi_j)\frac{(\chi-\chi_j)\chi_j}{a_j\chi}\mathop{\Delta\chi_j},
	\label{eq:convergence_discrete}
\end{equation}

\begin{figure}
\includegraphics[width=0.99\columnwidth]{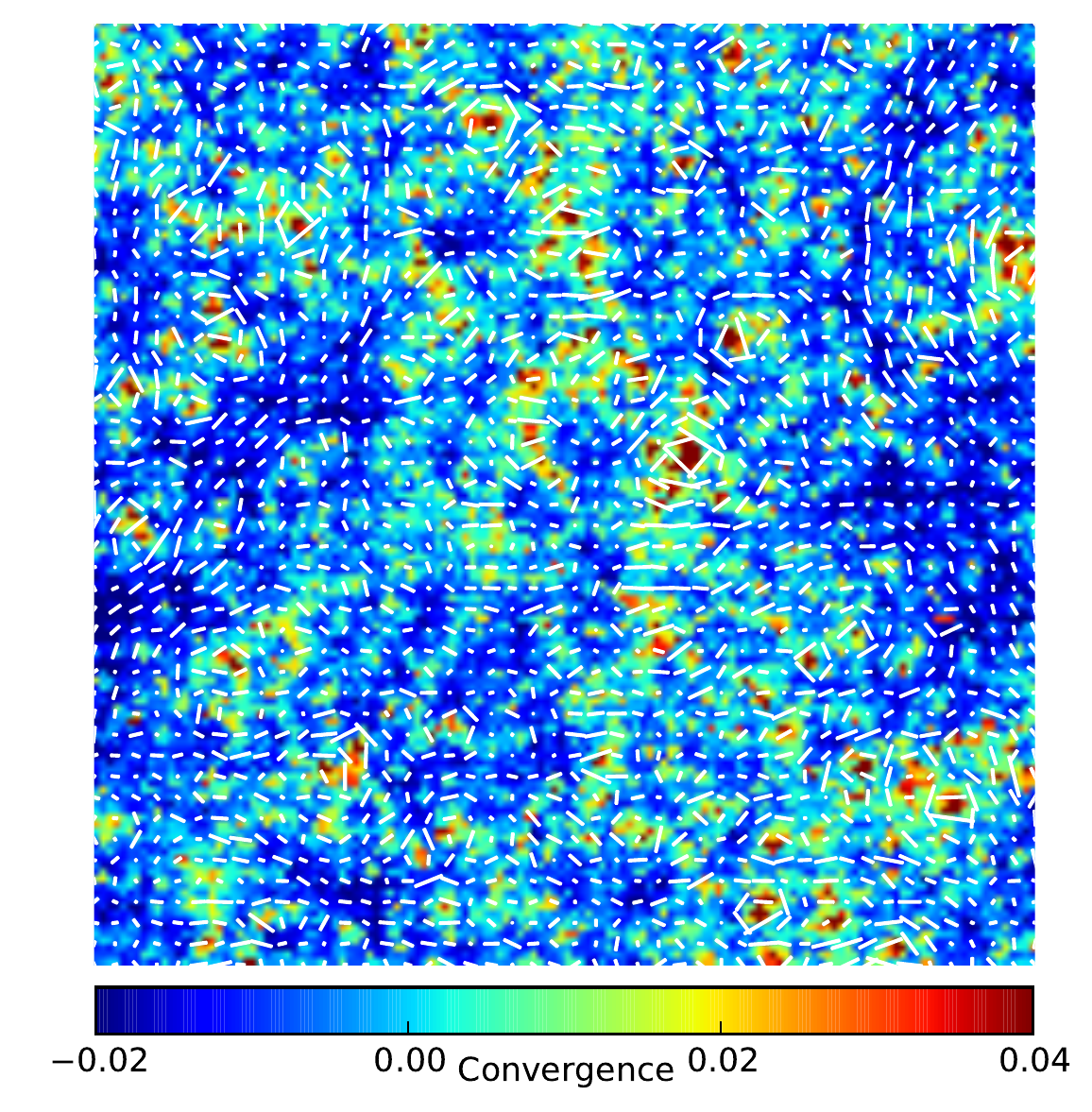}
\caption{Color map of the \icecola convergence field for a source redshift $z_s=1$ in a patch of $4\times4\, {\rm deg^2}$. The distribution features large voids with slightly negative values and concentrated peaks due to massive haloes or the superposition of several of them at intermediate distances. White ticks indicate the orientation and amplitude of the shear field (see Sec. \ref{sec:wl_shear}), and as expected for a field without $B$-modes they are preferentially oriented tangential to the peaks of the convergence field and point towards under-dense regions.}
\label{fig:kappa_shear_map}
\end{figure}

\noindent
where $\Delta\chi_j$ is the width of the radial bins.
The color map in Fig. \ref{fig:kappa_shear_map} shows the convergence field in a patch of $4\times4\, {\rm deg}^2$, which features the characteristic high density peaks of the cosmic web surrounded by large void regions.
The white ticks display the shear field, which we discuss in more detail below (see Sec. \ref{sec:wl_shear}).

It is useful to derive an expression for the convergence angular power spectrum starting from Eq. \ref{eq:convergence_born}. Using the Limber approximation \citep{Limber53} to compute the two-point statistics we can write \citep{Kaiser92,Kaiser98}:

\begin{equation}
	C_l^{\kappa}(\chi) = \frac{9H_0^4\Omega_m^2}{4c^4}\int_0^{\chi} P(k=\nicefrac{l}{\chi'},z)\frac{\chi-\chi'}{a^2\chi^2} \mathop{d\chi'},
	\label{eq:cl_kappa_prediction}
\end{equation}

\noindent
where $P(k,z)$ is the 3D matter power spectrum which can be derived from a theoretical model or measured from a numerical simulation. In the latter case, we can replace the integral by a sum over the redshifts where measurements for the matter power spectrum (produced on-the-fly after each time step) are available. The shot-noise contribution to the convergence angular power spectrum, $C_l^{\kappa,\, noise}$, can be computed using Eq. \ref{eq:cl_kappa_prediction}  by integrating the 3D shot-noise power spectrum,
$P(k,z)={1/\bar{n}}$, where $\bar{n}$ is the mean number density of particles,
weighted by the weak-lensing kernel.

\begin{figure}
\includegraphics[width=0.99\columnwidth]{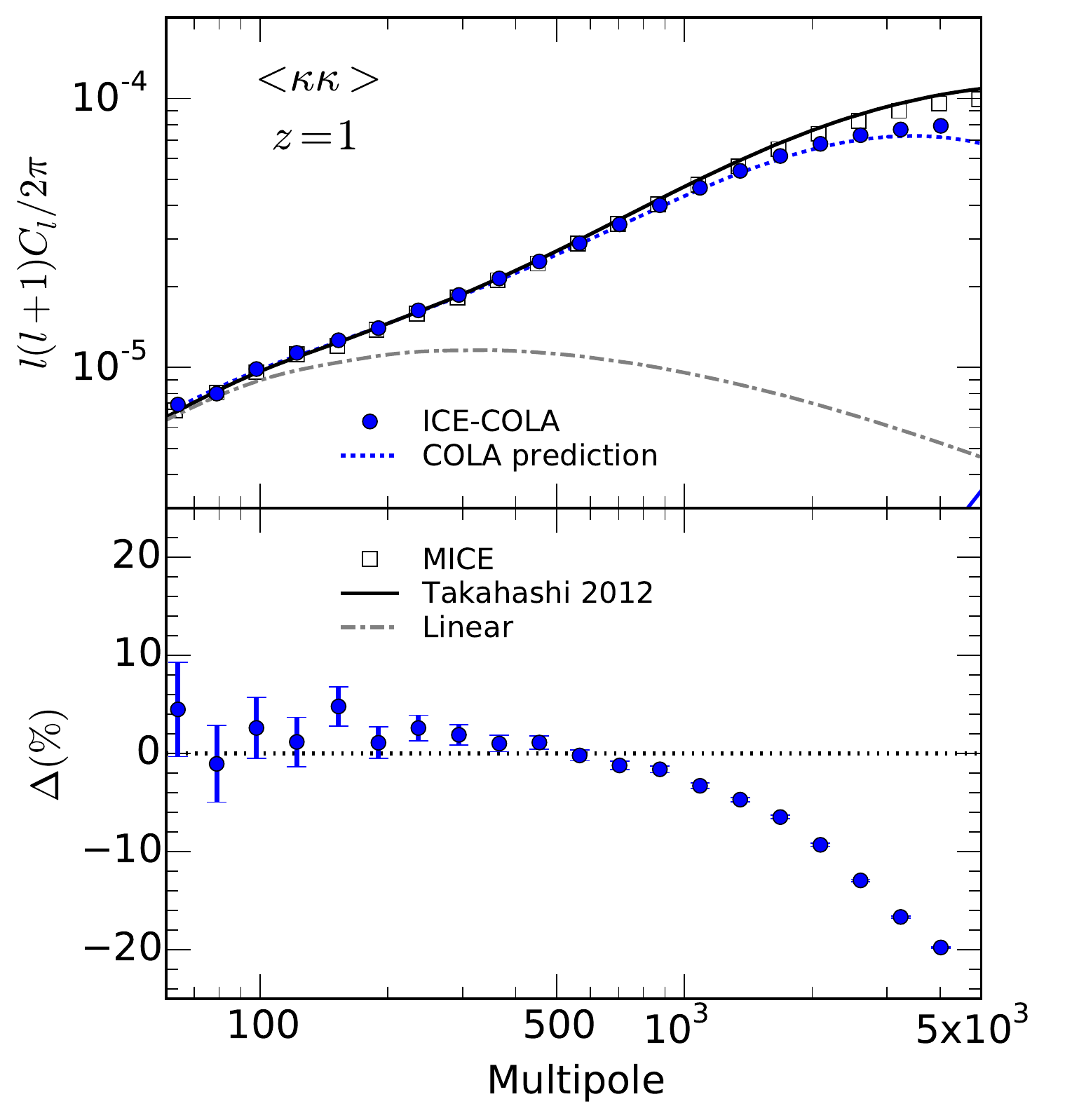}
\caption{Angular power spectrum for the convergence map at $z_s=1$. Different symbols represent: real measurements (circles), MICE-GC measurements (squares), theoretical prediction from equation \ref{eq:cl_kappa_prediction} and employing the \cola non-linear 3D matter power spectrum (dotted line), the non-linear prediction from revised Halofit (\citealt{Takahashi12} solid line) and the linear case (dot-dashed line). The shot noise contribution has been corrected, being below 10\% at all scales. \cola is in very good agreement with MICE-GC up to $l\sim1000$ and deviates thereafter to $10\%\, (20\%)$ at
$l=2\times10^3\,(4\times10^3)$. The theoretical prediction for \cola reproduces reasonably well the measurements (see the text why at small scales there is a small deviation).}
\label{fig:cl_kk}
\end{figure}

We show the measurements of the convergence angular power spectrum at a source redshift of $z_s=1$  in Fig. \ref{fig:cl_kk}. The curves display different theoretical predictions obtained with Eq. \ref{eq:cl_kappa_prediction}. In particular, the dotted curve uses the non-linear 3D matter power spectrum from \icecola. This prediction is in close agreement with the actual measurements from the maps, although it underestimates the signal at small scales\footnote{Note that at large scales, the deviations of the prediction using the \icecola non-linear matter power spectrum are due to sample variance.}.
This is because the measured 3D power spectrum used in equation  \ref{eq:cl_kappa_prediction} is available up to a maximum wavenumber. As a consequence, the prediction doesn't account for the contributions above that wavenumber, which are relevant in the limit of short distances $\chi'$ and high multipoles $l$, that correspond to high $k=l/\chi'$.
When compared to our fiducial MICE-GC reference measurements, \cola recovers the power accurately up to $l\sim10^3$ and declines thereafter. The difference is $\sim10 \,(20)\%$ at $l=2 \times10^3 (4\times10^3)$. Note that theses are highly non-linear scales, one decade above the transition scale to the non-linear regime (i.e., where the non-linear growth starts being significant, at $l\sim100$), and roughly correspond to $k\sim1\hompc$ at the distance where the lensing efficiency peaks.

\begin{figure}
\includegraphics[width=0.99\columnwidth]{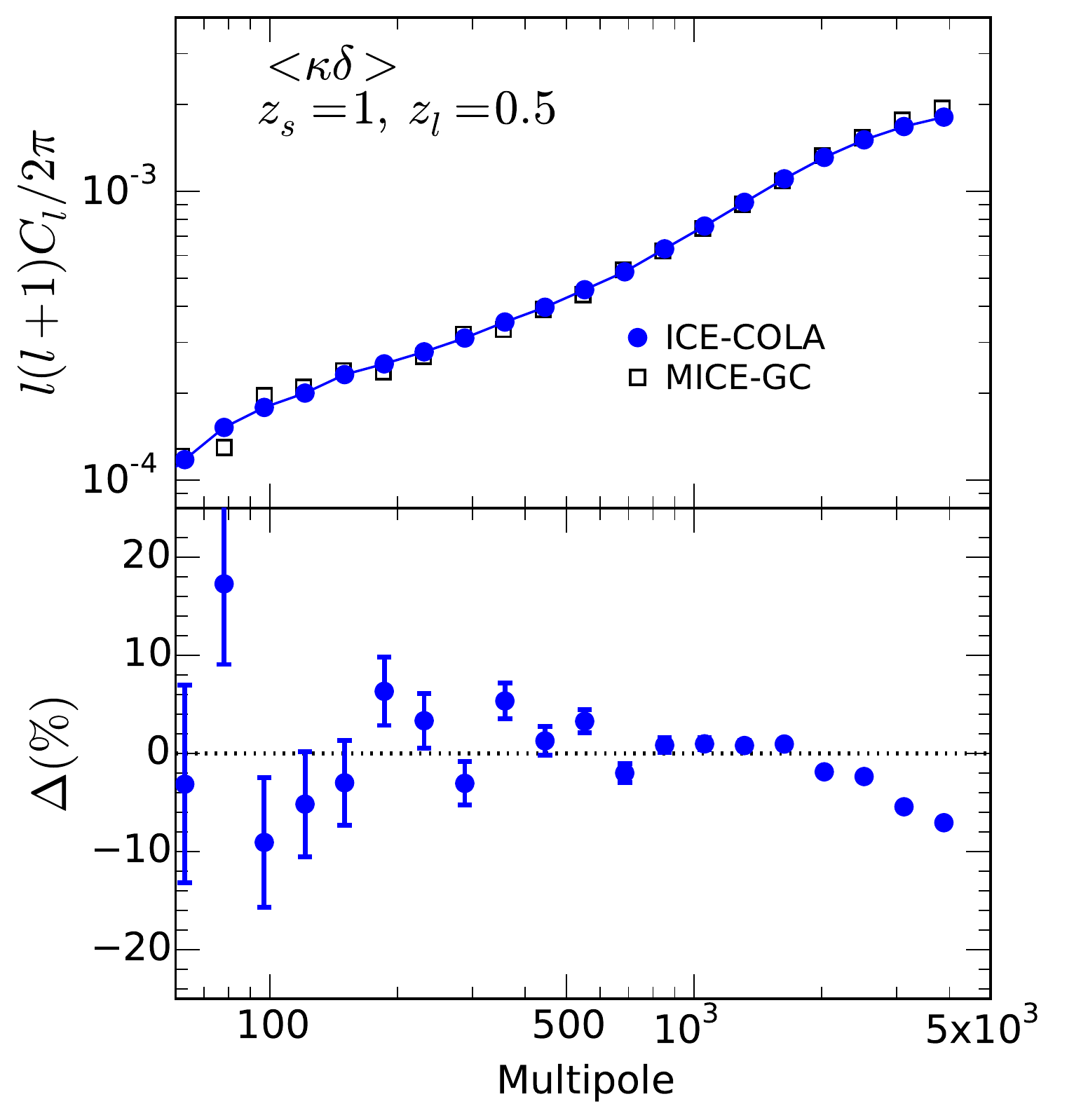}
\caption{Cross-power spectrum between the convergence field at $z_s=1$ and the density field at $z_l=0.5$ (with a width for the latter of $\Delta z = 0.1$). Deviations are found only at scales above $l=2000$, as in the case of the matter spectrum (see Fig. \ref{fig:cl_dd_m}).}
\label{fig:cl_kd}
\end{figure}

The convergence field is largely correlated with the intervening mass distribution due precisely to the lensing effect\footnote{This effect is closely related to what is known as galaxy-galaxy lensing}. This is shown in Fig. \ref{fig:cl_kd} for the cross-correlation between the convergence at $z_s=1$ and a lens sample of the matter density field in a lens plane at $z_l=0.5$ with width $\Delta z=0.1$. In this case, a good agreement is observed between \icecola and MICE-GC up to $l\sim2000$ and the deviations are within 10\% at most. Thus the cross-correlation with the matter field is better reproduced than the auto-power spectrum of the convergence. While the former measures the contribution to the convergence by the matter density field at the lens position, the latter has contributions also from high $k$-modes of structures at low redshift, which are more non-linear.
Therefore, at a given angular scale, the convergence auto-power spectrum has more non-linear contributions, hard to model in \cola, explaining why differences are larger in Fig \ref{fig:cl_kk} than in Fig. \ref{fig:cl_kd} and the latter resembles the behaviour seen in Fig. \ref{fig:cl_dd_m}.

\subsection{Shear maps}
\label{sec:wl_shear}

The shear field, $\gamma$, is a spin-2 tensor that in the complex notation can be expressed in terms of the polarization Stokes parameters ($\gamma_1,\gamma_2$): $\gamma=\gamma_1 + i \gamma_2$. In the all-sky formalism, the decomposition in harmonic space is written in terms of the spin-2 spherical harmonics $_{\pm2}Y_{lm}$ \citep{Hu00}

\begin{equation}
	\gamma_1(\bm\theta) \pm i \gamma_2(\bm\theta)
	= \sum_{lm} \hat{\gamma}_{lm}(\bm\theta)\, _{\pm2}Y_{lm}(\bm\theta).
	\label{eq:e-b_modes}
\end{equation}


The shear can be decomposed in the so-called $E$- and $B$-modes as $_{\pm}\hat{\gamma}_{lm}= \hat{\epsilon}_{lm} \pm i \hat{\beta}_{lm}$
In the weak gravitational lensing limit and in the Born approximation, the $B$-mode shear is zero, $\beta_{lm}=0$, due to the parity symmetry of the weak lensing field
(see e.g. \citealt{Cooray02b}). Then the shear is related to the convergence in harmonic space as (see \citealt{Hu00}):

\begin{equation}
 	\hat{\epsilon}_{lm}=-\sqrt{\frac{(l+2)(l-1)}{l(l+1)}}\hat{\kappa}_{lm}.
 	\label{eq:shear_e_mode}
 \end{equation}

Note that in the small angle limit ($l \gg 1$) the coefficient is just -1 and the power spectra of both quantities coincide.

Finally, the Stokes parameters of the shear field are given by (in the sign convention adopted by Healpix and in the absence of $B$-modes):

\begin{gather}
	\label{eq:shear_components1}
	\gamma_1 = - \sum_{lm} \hat{\epsilon}_{lm} X_{1,lm} \\
	\gamma_2 = \sum_{lm} i \hat{\epsilon}_{lm} X_{2,lm},
	\label{eq:shear_components2}
\end{gather}

\noindent
with $X_{{1\atop2},lm}$ defined as: $X_{{1\atop2},lm} = \nicefrac{(_2Y_{lm} \pm _{-2}Y_{lm})}{2}$.

In summary, starting from an all-sky convergence map we compute the shear by
\begin{enumerate*}
  \item transforming the map to harmonic space,
  \item computing the $E$-mode shear using equation (\ref{eq:shear_e_mode}),
  \item determining the two components in angular space of the shear by the inverse harmonic transforms given by equations (\ref{eq:shear_components1}) and (\ref{eq:shear_components2}) \footnote{The inverse harmonic transforms are performed using Healpy, \url{https://healpy.readthedocs.io/en/latest/}}.
\end{enumerate*}

Fig. \ref{fig:kappa_shear_map} displays the resulting shear map (white ticks), overlaid on the convergence field (scalar field, color coded). The shear is tangentially oriented to the convergence peaks, as expected for a cosmological weak lensing field (i.e, no $B$-mode, which would generate swirl or $45^{\circ}$ rotated patterns). An extensive validation of the shear is presented in the next section, where we also discuss  two-point clustering statistics for the light cone halo catalogues.

\section{Halo catalogues for weak lensing and clustering}
\label{sec:halo_weak_lensing_mocks}

Although we concentrate here on halo catalogues, we can think of them as a simple galaxy catalogues populated only with central galaxies that live in the center of dark-matter halos.  In the last step of the post-processing pipeline, we combine the two types of light cone outputs of the simulations: the halo catalogues and the weak lensing maps derived from the projected matter density field. In particular, the position of each halo determines the pixel in the 3D volume discretizaton of the light cone (as explained in Sec. \ref{sec:hp_maps}) that is used to assign the convergence and shear values.
For illustrative purposes, we generate two halo samples at well separated redshifts: the source sample at $z=1$  and the lens sample at $z=0.5$. Both catalogues are all-sky, have a radial width of $\Delta z=0.1$ and include objects with 100 or more particles (i.e., halos of mass larger than $3\times10^{12}\Msun$). This results in 6M and 11M objects for the lens and source samples, respectively, which correspond to number densities of 0.3 and $0.6\, {\rm arcmin^{-1}}$. The same criteria are used to draw the reference $N$-body samples from MICE-GC, except the latter were limited to only one octant of the sky.

In this section we use an additional \icecola simulation (with the same exact parameters and initial conditions as the default one) with the aim to distinguish the effects of the approximate dynamics of the simulation from the limited angular resolution employed in the Healpix sky maps. For that, we stored a light cone catalogue with particle information\footnote{The overhead due to writting particles in the catalogues was a 10\% of the running time for storing one octant of the light cone.} that we used to construct at post-processing a higher resolution version of the 2D projected matter density field\footnote{The light cone and therefore the catalogues generated are restricted to one octant for this simulation. However, this is sufficient to appropriately sample all the relevant scales discussed in this paper.}. We recall that the maps produced on-the-fly use the Healpix parameter $N_{side}=2048$ while for this test we use $N_{side}=4096$. In turn, MICE-GC catalogues were built with $N_{side}=8192$, but have checked that degrading them to $N_{side}=4096$ makes no difference on scales $\theta > 0.5\, {\rm arcmin}$, which is well within the range of scales discussed in this paper. Thus for all practical purposes there is no difference between the high-resolution maps of \icecola and MICE-GC.
The reference we quote the  corresponding pixel size for these resolutions, $\Delta\theta=\sqrt{3/\pi}3600/N_{side}=$ 1.72, 0.85 and 0.43 arcmin, for $N_{side}=2048,4096$ and $8192$ respectively.
Lastly, using these higher resolution Healpix maps we run the post-processing pipeline and produce the corresponding halo catalogues with weak lensing properties.


\subsection{Convergence auto-correlation}
\label{sec:convergence_2pcf_halo}

\begin{figure}
\includegraphics[width=0.99\columnwidth,trim={0 5cm 0 3cm},clip]{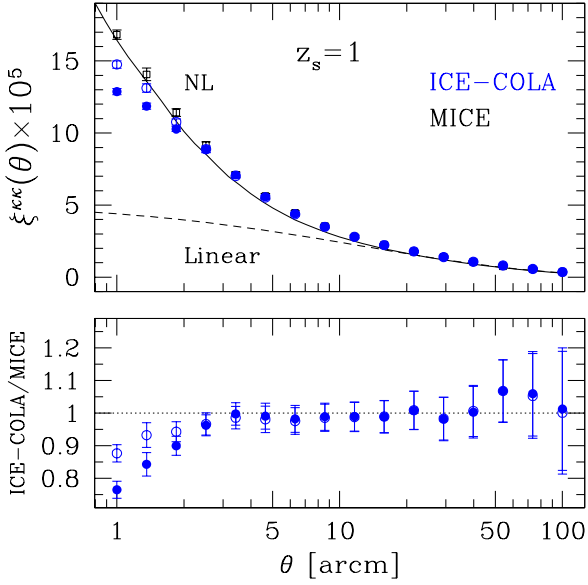}
\caption{Convergence correlation function for the source sample at $z_{s}=1$. For \icecola we show results for two samples that use weak-lensing properties sampled from Healpix maps at angular resolutions $N_{side}=2048$ and $4096$ (filled and open symbols, respectively). Error bars are computed by Jack-Kinfe resampling. The lower panel displays the ratio between \icecola and the MICE-GC simulation results. Non-linear predictions are computed using the revised Halofit (\citealt{Takahashi12}). It is shown that, below $\simeq 3$ arcmin, both \icecola samples underestimate the convergence power with respect to the exact N-body, although this effect is larger at a given scale when the lower angular resolution maps are used. On the other hand, above this scale, pixel effects have no impact.}
\label{fig:xikk_measured}
\end{figure}

The weak lensing maps describe the image distortions that an object would experiment at a given location of the sky. In practice, however, the maps are only sampled at those positions where there are sources (halos or galaxies). Nevertheless, this does not affect the 2-point weak lensing statistics of the halo lens sample, that are assigned from the weak lensing maps built from the underlying dark-matter field, and the measurement is also independent of the source sample galaxy bias.

Fig. \ref{fig:xikk_measured} shows the convergence correlation function in angular space for the source sample at $z_s=1.0$. In this work we measure correlation functions using the public code Athena\footnote{Athena is based on a two-dimensional tree, \url{http://www.cosmostat.org/software/athena/}.}. We find that \icecola reproduces the reference MICE-GC measurements at large scales and down to separations of $\sim 3$ arcmin. At smaller scales, \icecola shows a lack of power regardless of the angular resolution of the weak lensing maps, although this effect is smaller for the higher resolution maps
(e.g., $10\%$ at 2 or 1 arcmin for $N_{side}=2048$ or $4096$, respectively). Similarly as for the convergence power spectrum (see Fig. \ref{fig:cl_kk}), it is remarkable that the signal is well reproduced to separations one decade smaller than the scale where non-linearities arise. These results are in agreement with \cite{Heitmann10} (see their Fig. 1 and also sec. 3.3 in \citealt{MICEIII}), where it is investigated how a systematic effect in the matter power spectrum translates into a suppression of the shear correlation function at small scales. In the simulations used in this work, \cola recovers 50\% of the power in the matter power spectrum at $k\sim5\hompc$, which according to \cite{Heitmann10} should translate into a percent level accuracy down to scales of $\sim 3$ arcmin in the convergence angular correlation function, as we find.

The accuracy in the convergence two-point statistics is in rough agreement with the corresponding one from the angular power spectrum.
Since angular scales of $\sim2$ arcmin are approximately associated with multipoles $l \simeq \pi/ \theta \sim5000$ in the Limber limit, we should expect a similar accuracy in both statistics using this 1-to-1 relation in scales. However we find a better accuracy for the correlation function ($10\%$) than for the $C_l$'s ($20\%$).
A plausible explanation is that the approximate dynamics in \icecola introduce an error that resembles a scattering of the position of particles with a characteristic scale. Clustering measurements in configuration space will be correct at separations equal or larger than that scale. However, when transforming to harmonic space, small and large scales are mixed and errors are propagated to a wider range of scales (see e.g., \citealt{Monaco13}).


\subsection{Shear correlations}
\label{sec:shear_2pcf_halo}

The two components of the shear, that in the polar coordinates read $\gamma=\gamma_1+i\gamma_2$, can be expressed in a new and rotated base that is aligned with the separation vector between two points (where one point is associated with the shear map and the other with the positions of the objects in the lens sample, see e.g. \citealt{Kilbinger15}). The tangential $\gamma_t$ and the cross $\gamma_\times$ components of the shear are then defined as:
\begin{gather}
	\gamma_t      = - \Re(\gamma e^{-2i\phi}) \\
	\gamma_\times = - \Im(\gamma e^{-2i\phi}),
	\label{eq:gamma_tangential_cross}
\end{gather}
where $\phi$ is the polar angle of the separation vector.
The non-vanishing shear correlation functions are usually expressed in the two following combinations of the tangential and cross components (see e.g. \citealt{Bartelmann01})
\begin{figure}
  \includegraphics[width=0.99\columnwidth,trim={0 5cm 0 3cm},clip]{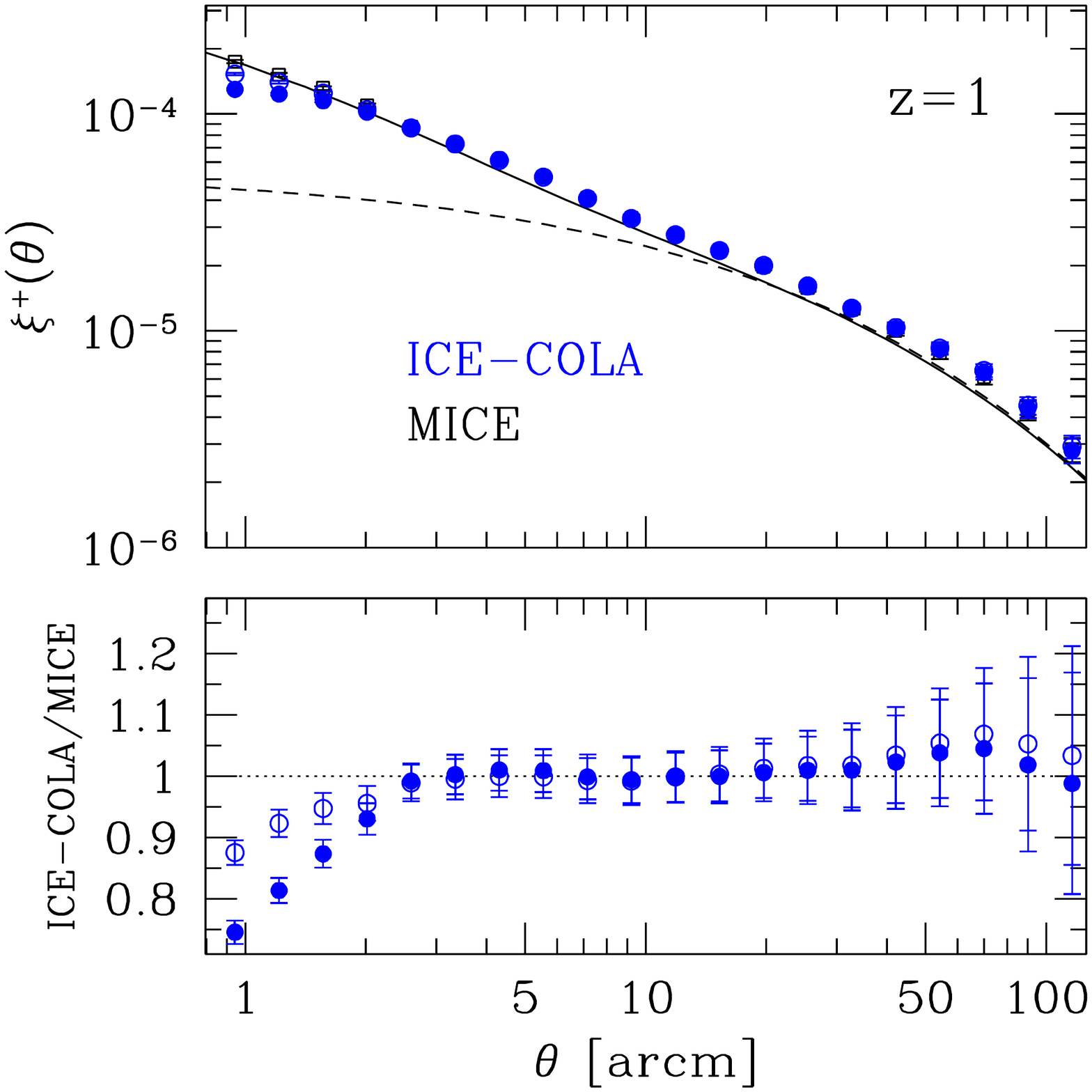}
\includegraphics[width=0.99\columnwidth,trim={0 5cm 0 3cm},clip]{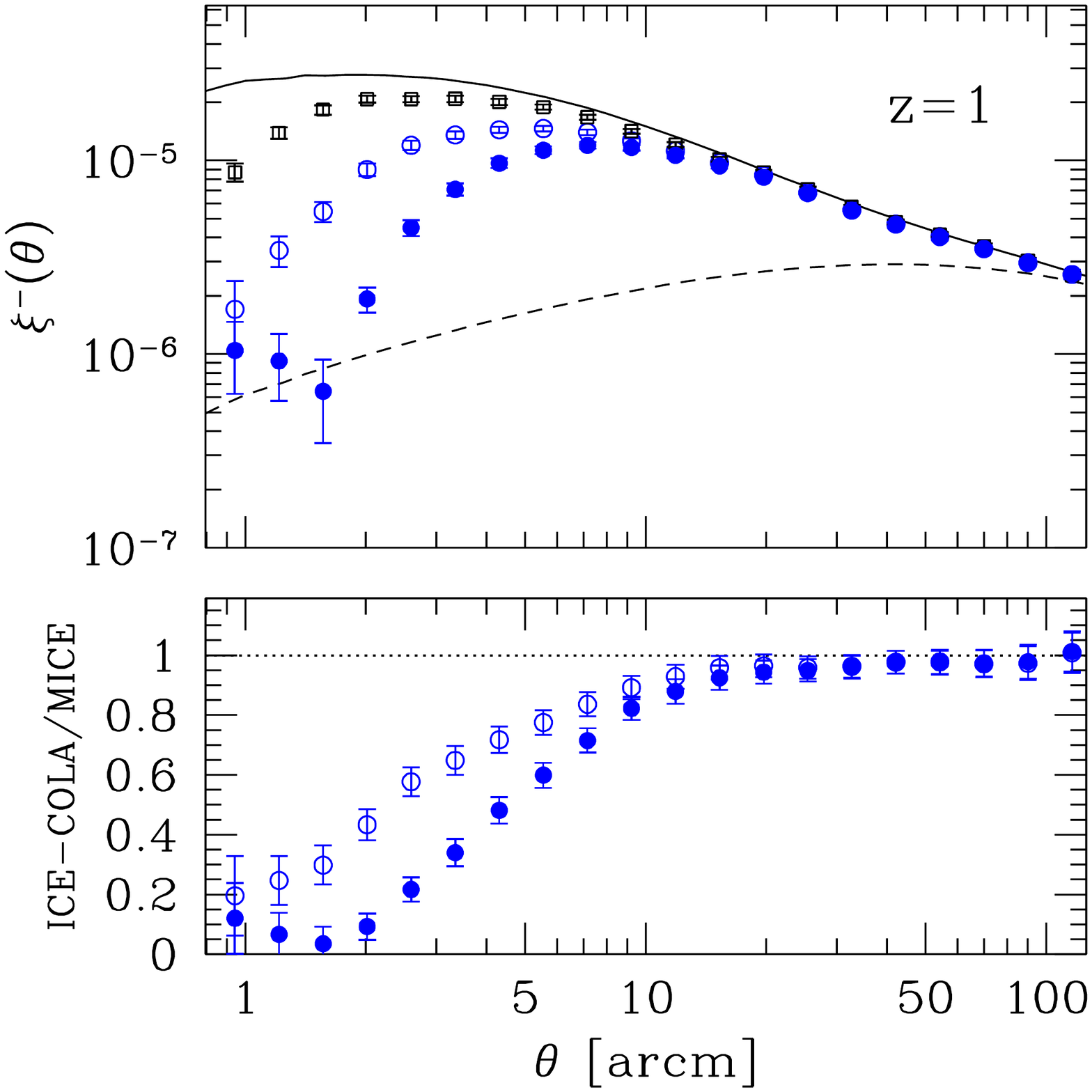}
\caption{Shear two-point correlation functions for the source sample (filled and open symbols corresponds Healpix
resolutions $N_{side}=2048$ and $4096$ respectively, as in Fig. \ref{fig:xikk_measured}). \icecola is percent accurate down to a scale that is one order of magnitude beyond the non-linear scale, i.e,  $\sim 2$ ($\sim20$) arcmin for the $\xi^+$ ($\xi^-$) component. As in Fig. \ref{fig:xikk_measured}, a higher angular resolution of the weak lensing maps does not change this characteristic scale, it only decreases the power loss on smaller angles.}
\label{fig:xipm_measured}
\end{figure}

\begin{equation}
	\xi_{\pm}(\theta) = <\gamma_t\gamma_t>(\theta)\, \pm <\gamma_\times\gamma_\times>(\theta).
\end{equation}

In the absence of shear $B$-mode and under the Limber approximation, the shear correlation functions can be expressed as (see \citealt{Bartelmann01}):

\begin{equation}
  \xi_\pm (\theta) = \frac{1}{2\pi} \int \mathop{dl} l J_{\nicefrac{0}{4}}(l\theta) C_l^{\kappa},
  \label{eq:shear_xi}
\end{equation}
where $J$ is the Bessel function of the first kind.

Fig. \ref{fig:xipm_measured} shows the shear correlation functions for the source sample at $z_s=1$. On large angles, \icecola is in good agreement both with the MICE-GC simulation and non-linear theory expectations.
The only caveat being the excess power in  $\xi^+$ for both simulations with respect to theory. One possible explanation is that we are using the small-angle limit for the prediction. Another is that the volume is not enough and both measurements are slightly biased (note that the points are very correlated). Nonetheless in this paper we are concerned about the agreement between \icecola and MICE-GC, and that is well achieved in Fig. \ref{fig:xipm_measured}.
On small scales, \icecola is percent accurate, as compared to MICE-GC, down to $\sim2$ and $\sim20$ arcmin for the $\xi^+$ and $\xi^-$ components, respectively. Such different ``characteristic accuracy scale'' (i.e,  the smallest angular scale at which \icecola matches N-body results)
comes from the fact that the minus component probes smaller scales of the density field, which are more non-linear, than the plus component. We can seen it in eq. \ref{eq:shear_xi}, since the $J_4$ Bessel function peaks at larger values of its argument than $J_0$.
In both cases \icecola resolves accurately scales down to one order of magnitude smaller than the non-linear scale (i.e., the scale where non-linearities become relevant), as we also found for the convergence correlation function. Also clear from Fig. \ref{fig:xipm_measured} The angular resolution of the lensing maps used does not alter this characteristic scale. But the coarser the angular resolution of the maps, the larger is the power loss at that characteristic scale, as shown in the lower panels of Fig. \ref{fig:xipm_measured}.

\begin{figure}
\includegraphics[width=0.99\columnwidth,trim={0 5cm 0 3cm},clip]{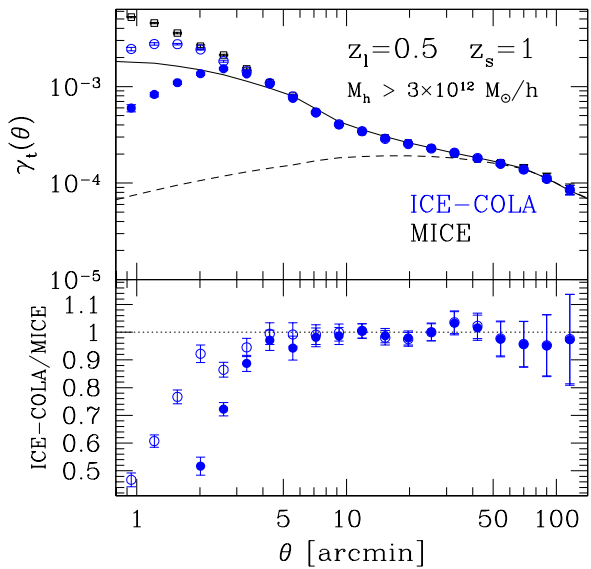}
\caption{Tangential shear for the source and lens samples at $z=1$ and 0.5 respectively of halos with $M >3 \times 10^{12} \Msun $ (symbols as in Fig. \ref{fig:xikk_measured}). Both \icecola samples reproduce accurately the signal down to about one decade below the non-linear scale (i.e, 4 arcmin in this case). The angular resolution of the weak lensing maps only affects the amplitude of the power loss at smaller scales, as in Fig. \ref{fig:xikk_measured} and \ref{fig:xipm_measured}.}
\label{fig:tshear_measured}
\end{figure}

Another important observable for weak lensing is the tangential shear around galaxies, also known as the galaxy-galaxy lensing.
The correlation function of the tangential shear $<\gamma_t\gamma_t>$ in the flat-sky limit, which is accurate enough for separations angles below few degrees (see \citealt{dePutter10}), it is given by

\begin{equation}
	\gamma_t(\theta)=\frac{1}{2\pi}\int J_2(l\theta)C_l^{\kappa\delta_h}l\mathop{dl},
\end{equation}
where $C_l^{\kappa\delta_h}$ is the cross-power spectrum between the source convergence and the lens number density.
Fig. \ref{fig:tshear_measured} shows the measurements of the tangential shear correlation function in \icecola for the source and lens samples at $z=1$ and $0.5$ respectively for halos with $M >3 \times 10^{12} \Msun $. There is a good agreement with MICE-GC for scales larger than 4 arcmin, regardless of the angular resolution employed for the weak lensing maps. Again, as for the previous lensing observables discussed, this matches the ``characteristic accuracy scale'' of \icecola, i.e, one decade below the non-linear scale. In the theoretical curves, the matter power spectrum has been multiplied by the linear bias factor that matches the clustering amplitude. The disagreement found on small scales with respect to theory predictions (which assume linear halo bias) may be attributed to a (positive) non-linear halo bias, that increases the clustering amplitude at small scales. The tangential shear depends not only on the weak lensing properties of the background sample but also on the clustering of the foreground one. To discriminate the impact of both on the tangential shear, we show in appendix \ref{sec:appendix} the angular correlation function of the lens halo sample. We conclude that at small scales, the errors from the halo bias seem to be sub-dominant in Fig \ref{fig:tshear_measured}, although halo auto-correlations start to depart from those in MICE-GC also at about $5$ arcmin.

\section{Conclusions}
\label{sec:conclusions}

In the advent of next generation of galaxy surveys, mock galaxy catalogues are key in the optimization of survey strategies and analysis pipelines. The increasingly large range of scales probed by weak lensing experiments put stringent requirements on the simulations needed to generate mock catalogues. We presented \icecola, a methodology to produce efficiently weak lensing maps and halo catalogues, both in the light cone geometry and generated on-the-fly. The speed-up thus gained with respect to standard $N$-body simulations is of 2-3 orders of magnitude and it requires a minimal post-processing of the catalogues compared to other techniques based on concatenation of snapshots to produce light cones. In this way, the amount of data to be stored is also greatly reduced.

We demonstrated the accuracy of the method by analysing the weak lensing properties of all-sky halo samples and compared them to the fiducial MICE-GC simulation. Running \icecola with the optimal parameter configuration described in \cite{Izard16}, we show that we can accurately model the two-point clustering of weak lensing observables one decade beyond the characteristic scale where the growth becomes non-linear. In particular, the convergence angular power spectrum is consistent with MICE-GC to multipoles of 1000, or 2-3 arcmin for the correlation function. For a fiducial case with sources at $z=1$ and lenses at $z=0.5$, we show that the shear correlation functions are accurate down to 2 and 20 arcmin for the $\xi^+$ and $\xi^-$ components respectively, whereas the tangential shear is accurately modeled down to 4 arcmin. These ``characteristic accuracy scale''  where \icecola breaks down is due to the approximate dynamics. We tested that our choice for the angular resolution of the weak lensing maps is balanced since improving it by a factor of two does not affect the scale where \icecola deviates, although it makes the power loss with respect to exact N-body simulations smaller.

Our implementation of \icecola opens the possibility of using approximate methods for the joint modeling of galaxy clustering and weak lensing observables and their covariance in the optimal exploitation of future galaxy surveys.


\section*{Acknowledgments}

We thank Linda Blot for her great help in some of the \icecola runs used in this paper and for helpful comments along the completion of the project. We are also grateful to the DEUS consortium (Jean-Michel Alimi, Pier-Stefano Corasaniti, and Yann Rasera) and Alina Kiessling for their help and infrastructure support.
AI was supported in part by Jet Propulsion Laboratory, California Institute of Technology, under a contract with the National Aeronautics and Space Administration. AI was also supported in part by NASA ROSES 13-ATP13-0019, NASA ROSES 14-MIRO-PROs-0064, NASA ROSES 12-EUCLID12-0004, and acknowledges support from the JAE program grant from the Spanish National Science Council (CSIC).
MC has been funded by AYA2013-44327 and AYA2015-71825-P. MC acknowledges support from the Ramon y Cajal MICINN program.
PF is funded by MINECO, project ESP2013-48274-C3-1-P, ESP2014-58384-C3-1-P, and ESP2015-66861-C3-1-R.
Simulations in this paper were run at the MareNostrum supercomputer - Barcelona Supercomputing Center  (BSC-CNS, \texttt{www.bsc.es}), through grants AECT-2015-1-0013, 2015-2-0011, 2016-2-0010, and 2016-3-0015. Funding for this project was partially provided by the Spanish Ministerio de Ciencia e Innovacion (MICINN).
Copyright 2017. All rights reserved.



\bibliographystyle{mnras}
\bibliography{biblist}


\appendix

\section{Halo correlation function}
\label{sec:appendix}

\begin{figure}
\includegraphics[trim={0 4.5cm 0 2cm},clip, width=0.99\columnwidth]{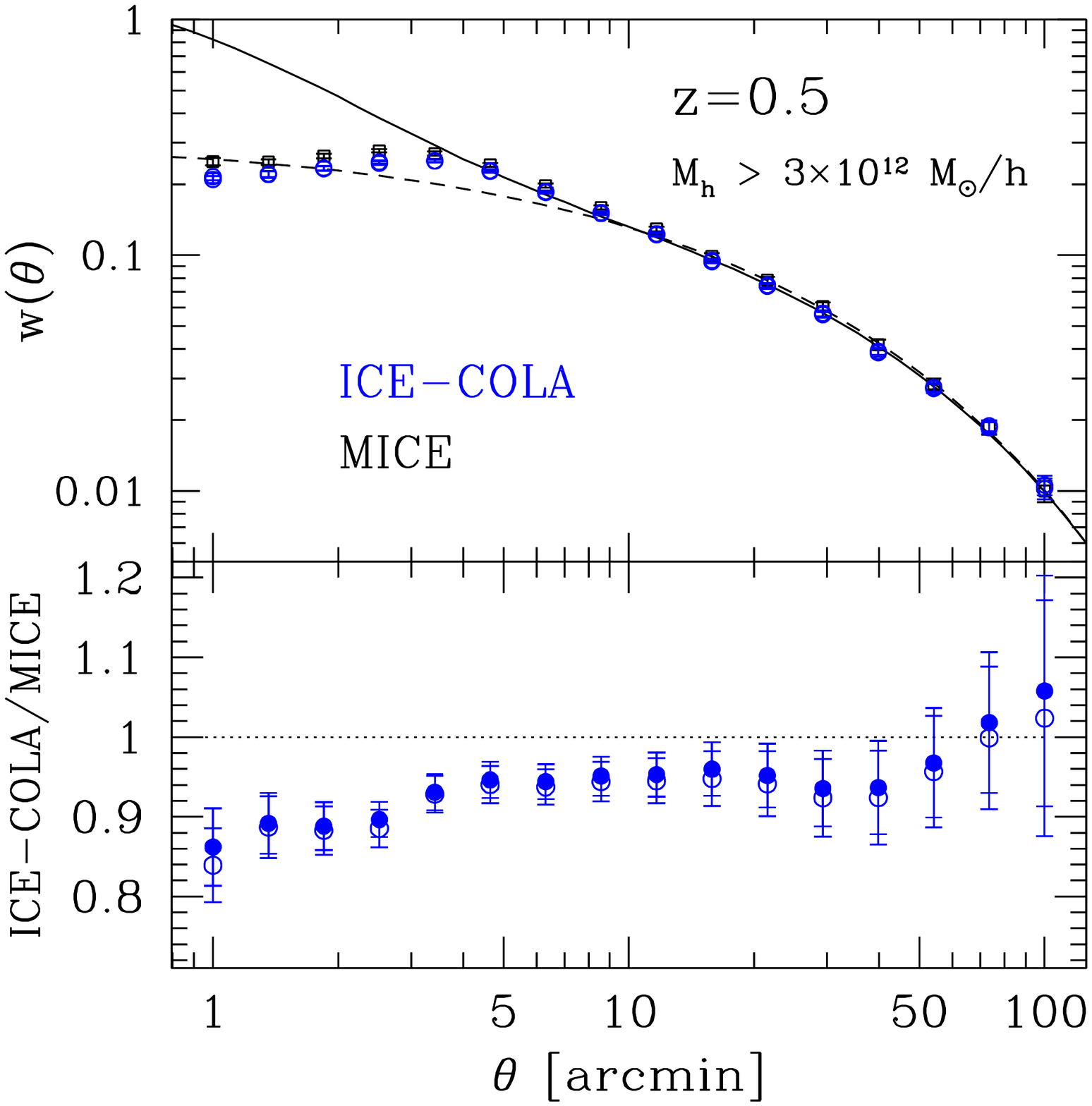}
\caption{Angular auto-correlation function for a foreground halo sample at $z=0.5$. The underestimation of the clustering amplitude at large scales is explained by a $2\%$ underestimation of the linear halo bias. At small scales, exclusion effects tend to reduce the measured power (see text for details).}
\label{fig:xi_halos}
\end{figure}

In this appendix, we discuss an observable that affects the tangential shear or galaxy-galaxy lensing: the halo clustering of a given lens population. In particular, the tangential shear contains information from both the weak lensing of the background sample and the bias of the foreground one. Therefore, for completeness, it is important to assess the accuracy on reproducing the clustering properties of the foreground sample.
Fig. \ref{fig:xi_halos} displays the halo angular correlation function. Simulation measurements deviate with respect to the non-linear prediction possibly due to both non-linear bias and exclusion effects, that reduce the power at scales close to the halo size. In fact, for the sample used, with halos of mass $M > 3\times10^{12}\Msun$ or corresponding size of $\sim 1 \mpcoh$, at $z=0.5$ they subtend an angle of $~\sim 4$ arcmin, in agreement with the scale at which we find possible exclusion effects in the measurements. The clustering amplitude at linear scales is $\sim5\%$ lower in \icecola, which is consistent with the $2\%$ underestimation of the halo linear bias (in turn due to an underestimation of halo masses, see \citealt{Izard16}).

\end{document}